\documentclass[11pt]{iopart} 
\usepackage[normalem]{ulem} 

\usepackage{bm,graphicx} 
\usepackage{mathrsfs,slashed,esint,booktabs} 
\usepackage{tikz}
\usetikzlibrary{shadings} 
\usepackage{iopams}  

\usepackage{amsfonts,bbm}

\usepackage{fancyhdr}

\newcommand{\eqref}[1]{(\ref{#1})}

\newcommand \Sphe {\mathbf S}

\newcommand \Arr \Acal  
\newcommand \Brr \Bcal  

\newcommand \Ars {\slashed \Acal}

\newcommand \mbb {\mathbbm m} 
\newcommand \notreH {\mathscr H}
\newcommand \notreM {\mathscr M}

\newcommand \seedg {g_{\textnormal{\scriptsize seed}}} 
\newcommand \seedm {m_{\textnormal{\scriptsize seed}}} 

\newcommand \seedh {h_{\textnormal{\scriptsize seed}}}

\newcommand \modu \infty

\newcommand \normal {{\textnormal{\scriptsize silh}}}  

\newcommand \Solu {\textnormal{\textbf{Sol}}}

\newcommand \cutoff c

\newcommand \expoP       {P}

\newcommand \expoPm {\underline P}

\newcommand \aweight \omega 
\newcommand \bweight \eta 

\newcommand \pweight p
\newcommand \qweight q

\newcommand \Deltaslash {\slashed\Delta{}}

\newcommand \nablaslash {\slashed\nabla{}}

\newcommand \xh {\widehat x}

\newcommand \unquad {\hspace{-1em}\ifmmode\mathopen{}\fi}

\newcommand \bfDiv {\textnormal{\textbf{Div}}}
 
\newcommand \Bcal 	{\mathcal B}

\newcommand \Acal {\mathcal A}

\newcommand \Mbf 	{\mathcal M} 

\newcommand \bse {\begin{subequations}}
\newcommand \ese {\end{subequations}}

\newcommand \la 	\langle
\newcommand \ra 	\rangle

\newcommand \Gcal 	{\mathcal G}

\newcommand \bei 	{\begin{itemize}}
\newcommand \eei 	{\end{itemize}}
\newcommand \del	 \partial

\newcommand \Fcal 	{\mathcal F}
\newcommand \Hcal 	{\mathcal H}

\newcommand \Mcal 	{\mathcal M}

\newcommand \RR 	{\mathbb R} 

\newcommand \eps 	\epsilon  
\newcommand \be 	{\begin{equation}\mathopen{}} 
\newcommand \ee 	{\end{equation}} 
\newcommand \lam 	\lambda 
  
\newcommand \Jbb {\mathbb J}  
\makeatletter
\let\orig@maketitle\maketitle
\let\orig@@maketitle\@maketitle
\let\orig@thanks\thanks
\let\orig@title\title
\let\orig@author\author
\let\orig@date\date
\let\orig@and\and
\newcounter{narticle}
\newcommand \newarticle {\clearpage\stepcounter{narticle}%
  \setcounter{section}{0}\setcounter{subsection}{0}\setcounter{equation}{0}%
  \addcontentsline{toc}{part}{Article \arabic{narticle}}%
  \gdef\theHsection{\arabic{narticle}-\arabic{section}}%
  \gdef\thesection{\fnsymbol{narticle}\arabic{section}}%
  \renewcommand\appendix{\gdef\theHsection{\arabic{narticle}-\Hy@AlphNoErr{section}}%
    \HyOrg@appendix\gdef\thesection{\fnsymbol{narticle}\Alph{section}}}%
  \let\thanks\orig@thanks
  \let\maketitle\orig@maketitle
  \let\@maketitle\orig@@maketitle
  \let\title\orig@title
  \let\author\orig@author
  \let\date\orig@date
  \let\and\orig@and
  \gdef\@date{\today}%
}
\makeatother

\newcommand{\ssAAUX}{\slashed{\Acal}}
\newcommand{\ssA}{\slashed{\ssAAUX}}
\declareslashed{}{/}{.1}{0}{\Acal}
\declareslashed{}{/}{.3}{0}{\ssAAUX}
\newcommand{\ssBAUX}{\slashed{\Bcal}}
\newcommand{\ssB}{\slashed{\ssBAUX}}
\declareslashed{}{/}{.1}{0}{\Bcal}
\declareslashed{}{/}{.3}{0}{\ssBAUX}
\newcommand{\ssFAUX}{\slashed{\Fcal}}

\declareslashed{}{/}{.1}{0}{F}
\declareslashed{}{/}{.3}{0}{\ssFAUX}
\newcommand{\ssrmAAUX}{\slashed{\mathrm{A}}}
\newcommand{\ssrmA}{\slashed{\ssrmAAUX}}
\declareslashed{}{/}{.1}{0}{\mathrm{A}}
\declareslashed{}{/}{.3}{0}{\ssrmAAUX}
\newcommand{\ssrmBAUX}{\slashed{\mathrm{B}}}

\declareslashed{}{/}{.1}{0}{\mathrm{B}}
\declareslashed{}{/}{.3}{0}{\ssrmBAUX}

\newcommand{\gdiff}{\gamma} 
\newcommand{\hdiff}{\eta} 

\newcommand{\pG}{p_{\mathrm{G}}}
\newcommand{\pA}{p_{\mathrm{A}}}

\renewcommand \textit {}

\makeatletter
\renewcommand{\section}{\@startsection{section}{1}{\z@}{-1.75ex \@plus -1ex \@minus -.2ex}{1.15ex \@plus .2ex}{\reset@font \normalsize \bfseries \raggedright}}
\renewcommand{\paragraph}{\@startsection{paragraph}{4}{\z@}{1.625ex \@plus 1ex \@minus .2ex}{-1em}{\reset@font \normalsize \itshape}}
\makeatother

\begin{document}

\pagestyle{fancy}
\fancyhf{}
\fancyhead[LE,RO]{Optimal shielding for Einstein gravity}
\fancyhead[LO,RE]{B. Le Floch and P.G. LeFloch}
\fancyfoot[C]{\thepage} 

\title{Optimal shielding for Einstein gravity}

\author{Bruno Le Floch$^1$ and Philippe G. LeFloch$^2$}

\address{$^1$ Laboratoire de Physique Th\'eorique et Hautes \'Energies, Centre National de la Recherche Scientifique \& Sorbonne Universit\'e, 4 Place Jussieu, 75252 Paris, France. 
\\
$^2$ Laboratoire Jacques-Louis Lions, Centre National de la Recherche Scientifique \& Sorbonne Universit\'e, 4 Place Jussieu, 75252 Paris, France. 
}
\ead{bruno@le-floch.fr, contact@philippelefloch.org}
\vspace{10pt}
\begin{indented}
\item[]February 2024.  Revised: April 2024
\end{indented}

\begin{abstract}
%
To construct asymptotically-Euclidean Einstein's initial data sets, we introduce the \textit{localized seed-to-solution method}, { which projects from approximate to exact solutions of the Einstein constraints.
The method enables us to glue together initial data sets in multiple asymptotically-conical regions, and in particular construct data sets that exhibit the gravity shielding phenomenon, specifically that are localized in a cone and exactly Euclidean outside of it.}
We achieve \textit{optimal shielding} in the sense that the metric and extrinsic curvature { are controlled at a \textit{super-harmonic rate,} regardless of how \textit{slowly they decay} (even} beyond the standard ADM formalism),
and the gluing domain can be a collection of \textit{arbitrarily narrow} nested cones.
We also uncover several notions of { independent interest:} { silhouette functions,} localized ADM modulator, and relative energy-momentum vector. { An axisymmetric example is provided numerically.}
\end{abstract}


\section{Shielding phenomena in Einstein gravity}
\label{section1} 

\paragraph{Shielding.}

We consider { three-dimensional initial data sets} for Einstein's vacuum field equations of general relativity, namely spacelike hypersurfaces embedded in a vacuum spacetime, modeled by a Ricci flat, { four-dimensional Lorentzian manifold.} Einstein's constraints form a set of four partial differential equations for the induced intrinsic geometry and extrinsic geometry. An extensive literature in physics and in mathematics is available on the construction of physically meaningful classes of solutions. In particular, one important strategy referred to as the \textit{variational method} was introduced by Corvino~\cite{Corvino-2000} and Corvino and Schoen~\cite{CorvinoSchoen} ---who built on Fischer and Marsden's work on deformations of the scalar curvature operator.
 
Motivated by the study of \textit{isolated} gravitational systems, we are interested here in asymptotically Euclidean, initial data sets and in the so-called \textit{anti-gravity} or \textit{shielding} phenomenon. Recall that, by the positive mass theorem, any vacuum space that coincides with the Euclidean space in the vicinity of an asymptotically Euclidean end is, in fact, globally isometric to the Euclidean space. In~\cite{CarlottoSchoen}, Carlotto and Schoen made a remarkable discovery by distinguishing between different angular directions at infinity. Namely, there exist non-trivial solutions to the constraints that, in a neighborhood of infinity, coincide with the Euclidean geometry in \textit{all angular directions except} within a conical domain with possibly small angle. Alternatively, the Euclidean and Schwarzschild solutions can be glued together at infinity across a conical domain. 
Chru\'sciel and Delay~\cite{ChruscielDelay-memoir,ChruscielDelay-2021,Delay} dealt with more general gluing domains and manifolds with hyperbolic ends. (Cf.~the reviews~\cite{Carlotto-Review,Chrusciel-bourbaki,GallowayMiaoSchoen} and Section~\ref{section6}.)   In this Letter, we follow this line of investigation.  Other recent important advances on the gluing problem include~\cite{AndersonCorvinoPasqualotto,AretakisCzimekRodnianski,CorvinoHuang,CzimekRodnianski,MaoTao} and further references cited throughout this text.  

\begin{figure}\centering\footnotesize\leavevmode
\raise60pt\rlap{(a)}\includegraphics[width= 0.3\textwidth]{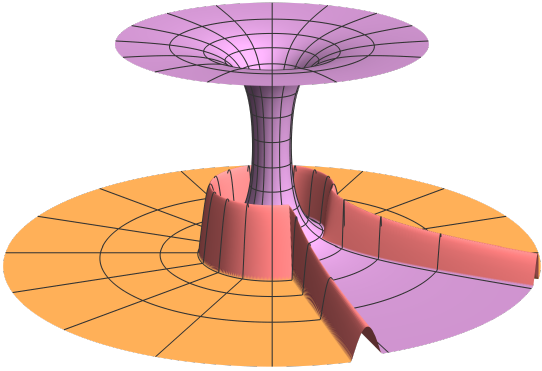}\hfill
\raise60pt\rlap{(b)}\includegraphics[width= 0.3\textwidth]{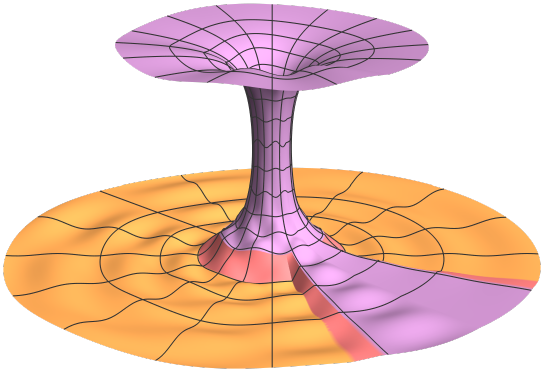}\hfill
\raise60pt\rlap{(c)}\includegraphics[width= 0.3\textwidth]{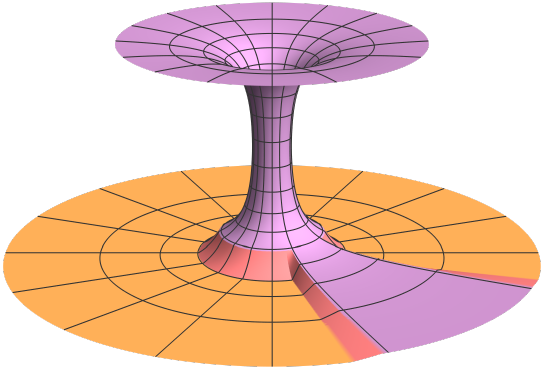}%
\caption{Euclidean metric (orange) glued to the Schwarzschild metric (purple) across a region that is conical at infinity.
(a) Exact sub-harmonic localization.
(b) Asymptotic super-harmonic localization.
(c) Exact super-harmonic localization.}
\label{figure---1} 
\end{figure}

\paragraph{Asymptotic shielding.}

The gluing scheme proposed by Carlotto and Schoen~\cite{CarlottoSchoen} allowed them to establish \textit{sub-harmonic} estimates within the localization region, { namely estimates in $r^{-1+ \eta}$} with respect to a radial coordinate $r$ at infinity, for arbitrarily small exponent $\eta>0$.  (C.f.~Figure~\ref{figure---1}(a).) Carlotto and Schoen conjectured that the localization with harmonic estimates should be achievable, that is, with $\eta=0$. Next, P.~LeFloch and Nguyen~\cite{LeFlochNguye3-preprint,LeFlochNguyen} proposed a different approach to the gluing problem and formulated what they called the \textit{asymptotic localization problem,} as opposed to the \textit{exact} version~\cite{CarlottoSchoen}. Therein, notion of \textit{seed-to-solution map} (cf.\ Section~\ref{section2}) was also introduced and sharp estimates at the \mbox{(super-)harmonic} level of decay were established, so that the gluing at \mbox{(super-)harmonic} rate was achieved in the sense that solutions enjoy the behavior required by the seed data in all angular directions at harmonic level at least, \textit{up to} contributions with faster radial decay. (C.f.~Figure~\ref{figure---1}(b).)  


\paragraph{Optimal shielding.}

 In contrast with \cite{CarlottoSchoen,LeFlochNguyen}, in the recent mathematical work~\cite{BLF--PLF} reported in this Letter, we achieve \textit{\mbox{(super-)}harmonic control} together with the exact localization sought by Carlotto and Schoen's conjecture ---as well as other ``optimal'' features discussed next. { This \textit{optimal localization} relies on the \textit{localized seed-to-solution projection,} as we call it, a tool we outline in Section~\ref{section2}.} Our method allows for the construction of broad classes of solutions which are of physical interest, and simultaneously provides an approximation scheme readily applicable for computational simulations. { It is based on a study of the linearized structure of the Einstein constraints using geometric analysis techniques, together with estimates on nonlinearities.}
Precisely, we construct classes of solutions that 
\begin{itemize}

\item[(i)] are defined by gluing within conical domains, possibly nested with narrow angles,  

\item[(ii)] { have any decay at infinity which can be {\it chosen} to be sub-harmonic or harmonic,} 

\item[(iii)] enjoy {\mbox{(super-)harmonic} decay} { in comparison with a prescribed seed data.} 

\end{itemize}

\noindent For a schematic illustration of the decay we refer to Figure~\ref{figure---1}(c). 
{ In our study, we  uncover several notions of physical and mathematical interest: silhouette functions (in kernels of asymptotic operators), localized ADM modulators, and relative energy-momentum vectors. Importantly, our work provides a complete and} direct validation of Carlotto and Schoen's conjecture. We now outline { the methods allowing} (i)--(iii) above, while referring to~\cite{BLF--PLF} for technical details and applications. 


\section{Localized seed-to-solution projection}
\label{section2}
   
\paragraph{Einstein constraints.}
 
{ We are interested in three-dimensional Riemannian manifolds} $(\Mbf, g, k)$ with (for simplicity) one asymptotic end, in which $g$~is a Riemannian metric and $k$~is a symmetric $(0,2)$-tensor field. Instead of~$k$ we work with the $(2,0)$-tensor 
$h = \big( k - \Tr_g(k) g\big)^{\sharp\sharp}$, 
where { $(\;)^\sharp$ denotes the map from covariant to contravariant tensors induced by~$g$}. The \textit{Hamiltonian and momentum constraints} in vacuum read   
\begin{equation}
\label{eq-aHM} 
\Hcal(g,h) 
= R_g + { {1  \over 2} }\bigl( \Tr_g h \bigr)^2 - | h |^2_g = 0, 
\qquad
\Mcal(g,h) = \bfDiv_g h = 0, 
\end{equation}
in which $R_g$ denotes the scalar curvature of $g$, and $\Tr_g h$ and $| h |_g$ denote the trace and norm of $h$, respectively, while $\bfDiv_g$ stands for the divergence operator.  


\paragraph{From seed data to initial data sets.}

We { shall parametrize} all solutions in the ``vicinity'' of a fixed \textit{localization data set} $(\Mbf, \Omega, g_0,h_0, r, \lambda)$, as we call it,
{
which specifies the underlying geometry of interest.
The gluing domain $\Omega \subset \Mbf$ is an asymptotically conical region, namely it is an exact cone in a preferred chart away from a compact region (see Figure~\ref{figure--3}, below).  It is equipped with a weight
\begin{equation} 
\omega_p = \lambda^{\expoP} r^{3/2-p}, 
\end{equation}
which behaves as a power of a radius function~$r$ on~$\Omega$, and  specifies a localization in angular directions based on a (large) power~$\expoP$ of a function $\lambda \geq 0$ vanishing linearly on the boundary of~$\Omega$ and independent of radius near the asymptotic end.
The ``reference'' data set $(g_0,h_0)$ (asymptotically Euclidean in the sense of~\eqref{equa-decayseed}, below) plays a secondary role in our construction and can be taken to be exactly Euclidean data $(\delta,0)$ near infinity.
We define a \textit{localized seed-to-solution projection} $\Solu_{p}$ that maps an approximate solution of the constraints $(\seedg,\seedh)$, close  to $(g_0,h_0)$, to an exact solution $(g,h)$ characterized  in \eqref{equa-deform}--\eqref{equa--221}, below.}
This standpoint extends the (non-localized) formulation in~\cite{LeFlochNguyen}.



\paragraph{Decay exponents: projection, geometry, and accuracy.}

It is essential  to clearly distinguish between the roles (and the ranges) of several pointwise decay exponents denoted by $p, \pG, \pA$ and satisfying 
\begin{equation}
{ p \in (0,1), \qquad } 
\pG \in (0, +\infty), 
\qquad 
\pA \geq \max(p,\pG) .
\end{equation}

\begin{itemize}
 
\item The \textit{projection exponent} $p$ for the map $\Solu_{p}$ arises in the variational formulation of the linearized Einstein operator. In short, we seek a deformation of the form
\begin{equation}\label{equa-deform} 
g = \seedg + \gdiff,
\qquad 
h = \seedh + \hdiff, 
\end{equation}
where, up to weights, $(\gdiff,\hdiff)$ belongs to the image of the adjoint $\rmd\Gcal_{(g_0,h_0)}^*$ of the linearization ---\textit{at the reference point} $(g_0, h_0)$--- of the Hamiltonian and momentum operators~\eqref{eq-aHM}. Specifically, there exist a scalar field $u$ and a vector field $Z$ so that 
\begin{equation}\label{equa--221} 
\gdiff = \omega_p^2 \, \rmd\Hcal_{(g_0,h_0)}^*[u,Z], 
\qquad \hdiff = \omega_{p+1}^2  \, \rmd\Mcal_{(g_0,h_0)}^*[u,Z].
\end{equation}

\item The \textit{geometry exponent} $\pG$ specifies the (possibly very low) decay of the seed metric and extrinsic curvature $(\seedg,\seedh)$ relative to Euclidean data $(\delta,0)$, namely
\begin{equation}\label{equa-decayseed} 
\seedg - \delta = \Or(r^{-\pG}), \qquad \seedh = \Or(r^{-\pG-1}).
\end{equation}

\item The \textit{accuracy exponent} $\pA$ describes how well the seed data obeys the constraints in the vicinity of the asymptotic end, in the sense that
\begin{equation}\label{HM-decay}
\Hcal(\seedg,\seedh) = \Or(\lambda^{\expoP} r^{ -\pA-2}), \quad
\Mcal(\seedg, \seedh) = \Or(\lambda^{\expoP} r^{ -\pA-2})
\end{equation}
and, in the case $\pA=1$, the Hamiltonian and momentum are integrable.

\end{itemize}

\noindent Rigorous results are stated in \textit{suitably weighted and localized} Lebesgue-H\"older norms (which we omit here) and require dealing with Killing initial data sets~\cite{Moncrief-1975}.  The above conditions { ensure that the projection map $\Solu_{p}$ exists, and that $(g-\seedg,rh-r\seedh)$ decays at least as $r^{-p_\star}$ for any exponent with $p_\star<1$ and $p_\star\leq \pA$, as}
discovered in~\cite{CarlottoSchoen} and slightly extended in~\cite{BLF--PLF}.


\section{Optimal shielding from within narrow gluing domains}
\label{section3} 

\paragraph{Harmonic stability.}

The { main result in~\cite{BLF--PLF} consists in \textit{sharp integral and pointwise estimates} (cf.~\eqref{equa-xjs3}, below) beyond the harmonic rate $r^{-1}$, provided the accuracy exponent $\pA>1$ is itself super-harmonic.  It involve a fourth exponent.}

\begin{itemize}

\item The \textit{sharp decay exponent} $ p_\star \in (1, \overline p_\star)$ with $p_\star\leq \pA$, controls how the difference $(g-\seedg, h-\seedh)$ between the seed data set and actual solution decays.

\end{itemize}

\noindent Here, $ \overline p_\star>1$ denotes an upper bound exponent which depends upon the localization function $\lambda$.
By examining an asymptotic version of the constraints at infinity (cf.\ Section~\ref{section4}), { we discover that $(g-\seedg,rh-r\seedh)$ contains an $r^{-1}$} harmonic contribution as part of the asymptotic structure of the solution at the asymptotic end, { which arises from the kernel of asymptotic operators on the sphere at infinity} (cf.\ Section~\ref{section5}).


\paragraph{Overview of the main theorem.}

We now outline our main conclusion. Consider a conical localization data set $(\Mbf, \Omega, g_0,h_0, r, \lambda)$ together with (projection, geometry, accuracy, { sharp decay}) exponents $(p, \pG,\pA,p_\star)$, as stated earlier.
{ Assume the localization function~$\lambda$, restricted to the sphere~$S^2$ at infinity, satisfies suitable stability conditions that amount to weighted Poincar\'e--Korn inequalities on its support (e.g.,\ the Hamiltonian functional \eqref{ssAalpha-quaform-0}, below, should be coercive).}

\begin{figure}
\centering
\includegraphics[width=0.33\textwidth, height=3.cm]{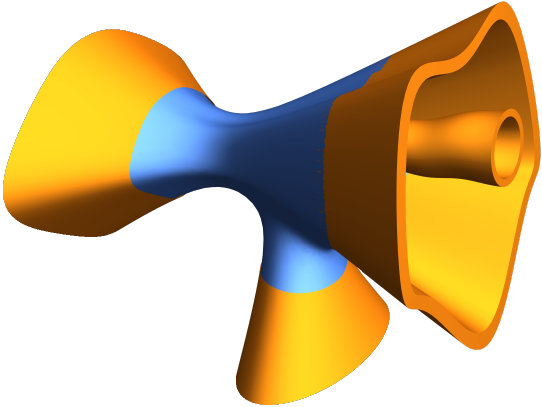}
\quad 
\begin{tikzpicture}[scale=.92]
  \draw[color=blue, domain=0:360, smooth, variable=\t, samples=200] plot ({cos(\t) + 0.05*sin(5*\t) + 0.03*sin(7*\t)}, {sin(\t) + 0.05*cos(5*\t) + 0.03*cos(7*\t)});
  \fill[color=blue, opacity=.2, domain=0:360, smooth, variable=\t, samples=200] plot ({cos(\t) + 0.05*sin(5*\t) + 0.03*sin(7*\t)}, {sin(\t) + 0.05*cos(5*\t) + 0.03*cos(7*\t)});
  \node at (0,0) {$\Omega$};
\begin{scope}[rotate=36]
    \draw[color=orange!80!black, fill=orange!80!black, fill opacity=.2] (2.3,0) ellipse(.5 and 1);
    \draw[color=orange!80!black, fill=white, fill opacity=.9] (2.3,0) ellipse(.43 and .93);
    \filldraw[color=orange!80!black, fill={orange!80!black!30!white}, fill opacity=.7]
    (2.1913043478260867, 0.9760845356801586)
    -- (0.657391304347826, 0.2928253607040476)
    arc (102.55585779858599:257.444142201414:.15 and .3)
    -- (2.1913043478260867, -0.9760845356801586)
    arc (257.444142201414:102.55585779858599:.5 and 1);
    \draw[color=orange!80!black, very thin] (.69,0) ellipse(.15 and .3);
    \draw[color=orange!80!black, very thin] (.69,0) ellipse(.129 and .279);
    \node at (3.05,0) {$\Omega_R$};
\end{scope}
\end{tikzpicture}
\quad
\begin{picture}(0,0)\footnotesize
  \put(23,47){$\lambda$}
  \put(51,47){$\sigma$}
  \put(6,-7){$\theta_1$}
  \put(137,-7){$\theta_2$}
\end{picture}%
\includegraphics[height=3.cm]{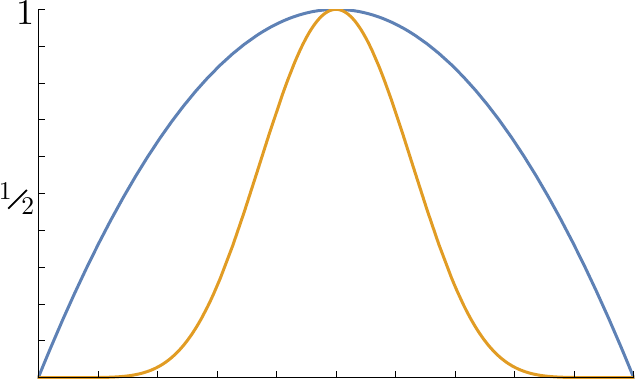}
\caption{
Left: {gluing domain with four asymptotic ends (orange) diffeomorphic to thin truncated cones, and connected by an interior region (blue).} 
Middle: schematic representation of a gluing domain $\Omega$. 
Right: localization functions $\lambda$ and $\sigma$ (Section~\ref{section5bis}).}
\label{figure--3}
\end{figure}
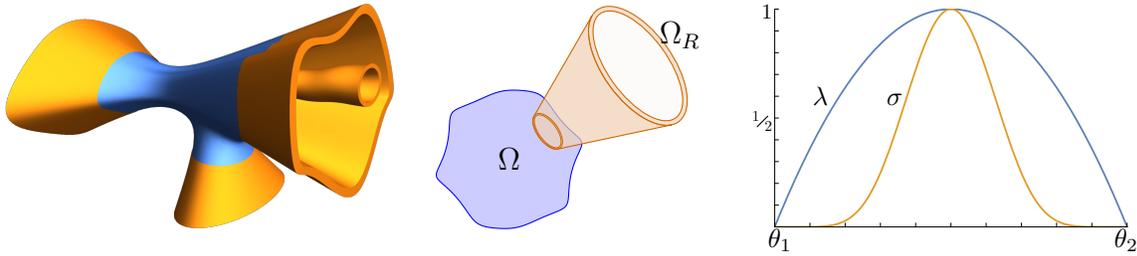

\begin{itemize}

\item For any localized seed data set $(\seedg, \seedh)$ that is sufficiently close to $(g_0, h_0)$, there exists a solution $(g,h)$ to the  constraints defined in \eqref{equa-deform}--\eqref{equa--221} by variational projection of $(\seedg, \seedh)$ which enjoys the pointwise decay (for some large $\expoPm$ with $\expoPm<\expoP$)
\begin{equation}
\label{equa-xjs3} 
\fl
g = \seedg +  \kappa \, g^\modu 
+ \Or(\lambda^{\expoPm} r^{-p_\star}), \qquad 
h = \seedh +  \kappa \, h^\modu + \Or(\lambda^{\expoPm} r^{-p_\star-1}). 
\end{equation}

\item In~\eqref{equa-xjs3}, the modulated seed data set $(\seedg + \kappa \, g^\modu , \seedh +  \kappa \, h^\modu)$, 
as we call it,  is defined in \eqref{equa-2p82}, below, and coincides with the prescribed seed data set $(\seedg, \seedh)$ { {\it up to ``localized ADM modulators''} described in} Section~\ref{section5}, below.

\item 
 Importantly, { the stability conditions} required on the function $\lambda$ allows for gluing domains that are \textit{nested and/or arbitrarily narrow in one direction,} as illustrated by Figure~\ref{figure--3} {---in which we include several asymptotic ends (as is covered in~\cite{BLF--PLF}).}
 
\end{itemize}
\noindent We refer to these results as the {\it optimal shielding theory,} since we allow for solutions that are arbitrarily localized, are generated from data with arbitrarily slow decay, and yet enjoys estimates at the super-harmonic level of decay. { A full statement is available in~\cite[Theorem 1.1]{BLF--PLF} and we now present several key notions of our method.}

\section{ Harmonic constraint operators and stability conditions}
\label{section4}  

\paragraph{Notation.}

{ 
After a suitable reduction (not discussed here, based on treating nonlinearities as asymptotic perturbations),
we are led to consider a single truncated cone $\Omega_R$ of the Euclidean space $(\RR^3, \delta)$, defined by its support $\Lambda \subset S^{2}$ and minimum radius $R>0$, and consisting of points $x\in\RR^3$ with $r\geq R$ and $x/r\in\Lambda$, where $r=|x|$.
The weight $\lambda=\lambda(x/r)$ vanishes linearly near the boundary of its support~$\Lambda$, and the weight $\omega_{p} = \bigl(\lambda(x/r)\bigr)^{\expoP}r^{3/2-p}$ involves exponents $p \in (0, 1)$ and a sufficiently large~$\expoP$.
}%
The weighted Sobolev space $H^2_{- \expoP}(\Lambda)$ (used in \eqref{equa-stable-H-414}, below) is defined with respect to the measure $\lambda^{2 \expoP} \rmd\xh$, where $\rmd\xh$ is the standard volume form of the two-dimensional unit sphere $S^2$. The weighted average of a function $f\colon \Lambda \to \RR$ is 
\begin{equation}\label{equa-average} 
\la f \ra = \fint_{\Lambda} f \, \lambda^{2 \expoP} \rmd\xh 
= {1 \over |\Lambda|_\lambda} \int_{\Lambda} f \, \lambda^{2 \expoP} \rmd\xh,  
\qquad  
|\Lambda|_\lambda = \int_{\Lambda} \, \lambda^{2 \expoP} \rmd\xh. \qquad
\end{equation}


\paragraph{A decomposition of the Einstein constraints.}

The asymptotic properties of the seed-to-solution map are controlled by the linearization of the constraints around Euclidean data, evaluated on the deformation $(\gdiff,\hdiff)$ given in~\eqref{equa--221}.
This leads to fourth- and second-order (squared) localized Hamiltonian and momentum operators
\begin{equation}\label{equa:acalew0}
\eqalign{\notreH[u] = \omega_{p}^{-2} \, \Bigl( 2 \, \Delta(\omega_{p}^2 \, \Delta u)
+ \del_i\del_j(\omega_{p}^2) \del_i\del_j u - \Delta(\omega_{p}^2) \Delta u
\Bigr), \\
\notreM[Z]^i = - \frac{1}{2} \, (\Delta Z^i + \del_j \del_i Z^j) - (\del_j \log \omega_{p+1}) \, (\del_j Z^i + \del_i Z^j),}
\end{equation}
in which $u\colon \Omega_R \to \RR$ is scalar-valued and $Z\colon \Omega_R \to \RR^3$ is vector-valued, and where repeated indices are implicitly summed.
{
We concentrate here for simplicity on the Hamiltonian operator.
To study the \mbox{(super-)}harmonic decay of solutions, we split it as $r^4 \notreH[u] = \Arr[u] + \Ars[u] + \ssA[u]$,
where $\Arr[u]=2r\del_r(r\del_r+2-2p)(r^2\del_r^2+(3-2p)r\del_r-2)u$ contains only radial derivatives, $\Ars$ mixes radial and angular derivatives,
while $\ssA$} is defined by plugging functions $\nu \, r^{-2( 1-p)}$ (corresponding to harmonic terms in the metric):
\begin{equation}\label{equa-asymptoOper-reapeat} \fl
\ssA{}[\nu]  = r^{3(1-p)} \notreH[\nu \, r^{-2(1-p)}]
= \lambda^{-2\expoP} \Bigl( \Deltaslash(\lambda^{2\expoP}\Deltaslash\nu) + \nablaslash^2\cdot(\lambda^{2\expoP}\nablaslash^2\nu)
+ \mbox{lower order} \Bigr) ,
\end{equation}
where $\nablaslash$, $\nablaslash^2$, and~$\Deltaslash$ denote the covariant derivative, Hessian, and Laplacian on~$S^{2}$. 
This is an elliptic but non-self-adjoint operator due to two-derivative terms.
Integrating it against $\nu\,\lambda^{2 \expoP} \rmd\xh$ yields the quadratic functional
\begin{equation}\label{ssAalpha-quaform-0} 
\ssrmA[\nu]
= \fint_{\Lambda} \Bigl( (\Deltaslash\nu)^2
+ |\nablaslash^2\nu|^2
+ (6-4p) |\nablaslash\nu|^2 -  4p( 1-p) \,\nu \Deltaslash\nu
\Bigr) \lambda^{2 \expoP} \rmd\xh.
\end{equation} 


\paragraph{Stability conditions on the localization function.}

We uncover conditions to be imposed on the function $\lambda$ in order to establish \mbox{(super-)}harmonic estimates.
These stability conditions are \textit{weighted { Poincar\'e--Korn} inequalities} on the { support~$\Lambda$.} For instance, our \textit{harmonic stability condition} for the Hamiltonian reads 
\begin{equation}\label{equa-stable-H-414}
  \ssrmA[\nu] \gtrsim \| \nu \|_{H^2_{- \expoP}(\Lambda)}^2
  \qquad \mbox{for } \nu \in H^2_{- \expoP}(\Lambda) 
  \mbox{ with } 
   \la\nu\ra = 0.
\end{equation} 
In~\cite{BLF--PLF}, we { deduce from this condition} that  the kernel $\ker(\ssrmA)$ is of dimension~$1$ and we select an element~$\nu^{\normal}$, dubbed the \textit{silhouette function}, normalized by the condition $ \la \Deltaslash\nu^{\normal} - 4(1-p) \, \nu^{\normal} \ra = 8 \pi/ |\Lambda|_\lambda$. For the asymptotic operator~$\ssB$ associated with the momentum $\notreM$, we arrive at a kernel with dimension $n$, and normalized kernel elements $\xi^{\normal(j)} \in \ker\big( \ssB{} \big)$ (for $j=1,2,3$) referred to as the silhouette vector fields. 


\section{Localized ADM modulators and relative energy-momentum vectors}
\label{section5} 

\paragraph{Modulation at infinity.}

The seed-to-solution map generates harmonic terms, which we { are now ready to} describe.
Starting from the seed data set $(\seedg, \seedh)$ we introduce the \textit{modulated seed data set} $(\seedg +  \kappa \, g^\modu,  \seedh +  \kappa \, h^\modu)$ used in~\eqref{equa-xjs3}, where { $\kappa$ interpolates smoothly from $0$ in a (large) compact set to $1$ in the conical part of~$\Omega$}, and the \textit{localized modulator} $(g^\modu, h^\modu)$ is parametrized by a spacetime vector $(m^\modu, J^\modu)$:
\begin{equation}\label{equa-2p82}
\eqalign{
g^\modu_{ij}
=
\lambda^{2\expoP} r^{3-2p} \bigl(\del_i \del_j  u^\modu- \delta_{ij} \Delta  u^\modu \bigr), 
\qquad 
& u^\modu = m^\modu \nu^\normal(x/r) r^{-2+2p}, 
\\
h^\modu_{ij}
= - {1\over 2} \, \lambda ^{2\expoP} r^{1-2p} \bigl(\del_i Z^\modu_j + \del_j Z^\modu_i \bigr) ,
\quad
& Z^\modu = J_k^\modu \xi^{\normal(k)}(x/r) r^{-2+2p} .
}
\end{equation}

{

\paragraph{Standard ADM invariants.}

The Arnowitt-Deser-Misner (ADM) energy and momentum $\mbb(g)$ and $\Jbb(h)$ are standard invariants of asymptotically Euclidean initial data sets in general relativity, which are well-defined for the seed and solution for sufficiently strong decay $\pG>1/2$ and $\pA\geq 1$.
In that case, the spacetime vector $(m^\modu, J^\modu)$ arising in \eqref{equa-2p82} is actually the difference of ADM four-momenta of $(g,h)$ and $(\seedg,\seedh)$.
However, our theory allows for very slow decay of the metric ($r^{-\pG}$ for any $\pG>0$) and extrinsic curvature, in which case the ADM invariants are ill-defined.
Nevertheless, their difference can remain a well-defined notion of {\it relative invariant,} as follows.}


\paragraph{Relative ADM invariants.}

Given two { asymptotically Euclidean} pairs $(g,h)$ and $(g',h')$ of symmetric two-tensors on the same manifold, the \textit{relative energy} and the \textit{relative momentum vector} are defined (at each end, whenever the limits exist) as 
\begin{equation}\label{equa-def-mass-momentum}
\eqalign{
\fl \mbb(g-g') 
& = {1 \over 4 \, |S^{2}|} \lim_{r \to + \infty} 
r^{2} \int_{S^{2}} 
\sum_{i,j = 1}^3 {x_j \over r} \, \bigl( (g-g')_{ij,i} - (g-g')_{ii,j} \bigr) \Big|_{|x|=r} \rmd\xh,
\\
\fl \Jbb(h-h')_j
&=  {1 \over 2 \, |S^{2}|} \lim_{r \to + \infty} 
r^{2}  \int_{S^{2}} \sum_{1 \leq k \leq 3}{x_k \over r} \,  (h-h')_{jk} \Big|_{|x|=r} \rmd\xh. 
}
\end{equation}
{ While we have $\mbb(g-g') = \mbb(g) - \mbb(g')$ provided both masses are finite, the relative mass may be finite even when masses of $g$ and~$g'$ are infinite (and likewise for momenta): our definition only requires that the \textit{difference} has sufficient decay.
In detail, if the (nonlinear) constraints decay as $\Hcal(g,h),\Mcal(g,h)=O(r^{-\pA-2})$ and likewise for $(g',h')$, the data sets agree at a rate $r^{-a}$, and decay to $(\delta,0)$ at a rate $r^{-\pG}$ with $\pA\geq 1$, $a>0$, and $a+\pG>1$, then relative invariants are well-defined.
For $(g,h)$ and $(\seedg,\seedh)$ the conditions reduce simply to $\pA\geq 1$ as the sub-harmonic estimates discussed below~\eqref{HM-decay} allow to take $a$ arbitrarily close to~$1$.
%
In turn, our choice of normalization of the kernel elements ensures that the energy-momentum vector $(m^\modu, J^\modu)$ matches the relative invariant of the prescribed data set and actual solution, that is,
\begin{equation}
m^\modu = \mbb(g - \seedg),
\qquad
J^\modu = \Jbb(h - \seedh).
\end{equation}
}


\section{A numerical example: harmonic behavior at infinity}
\label{section5bis} 

\paragraph{Numerical investigations.}

The projection map $\Solu_{p}$ can be computed numerically, as our framework naturally leads to a suitable stable boundary value problem for nonlinear elliptic equations posed in a domain with singular behavior at the boundary, and our methods (stability inequalities, energy functionals, etc.\@) are amenable to numerical analysis. For illustration, we thus supplement our theoretical work~\cite{BLF--PLF} by  studying the Hamiltonian at infinity in the time-symmetric initial data sets $h \equiv 0$ with axisymmetry.  We focus on the key challenge overcome in the present work for reaching the harmonic decay, namely the analysis of the harmonic Hamiltonian operator~$\ssA$ on the sphere at infinity.  We do not attempt to review recent numerical developments such as \cite{BFR}. 

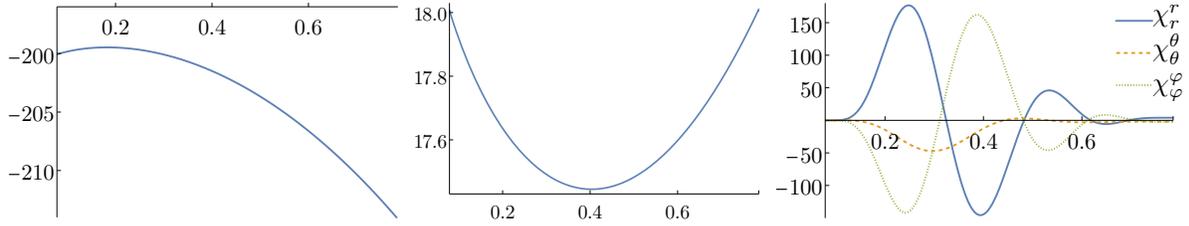
\begin{figure}\centering\footnotesize\leavevmode
\includegraphics[height=2.9cm]{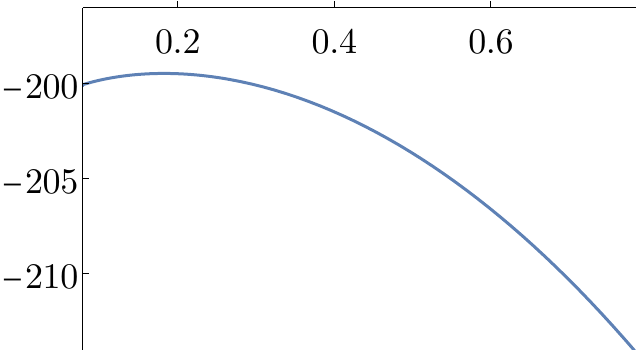}
\hfil
\includegraphics[height=2.9cm]{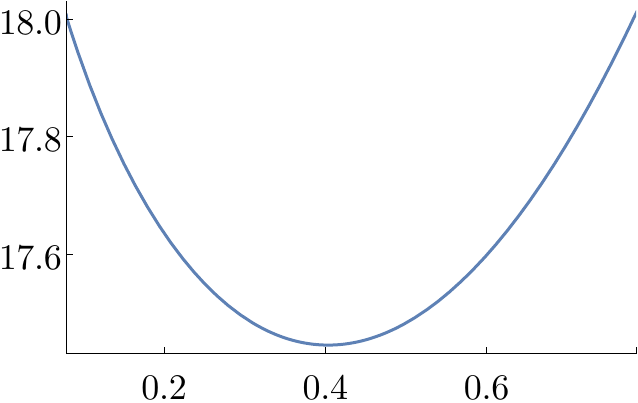}
\hfil
\begin{picture}(0,0)\footnotesize
  \put(143,75){$\chi_r^r$}
  \put(143,62){$\chi_\theta^\theta$}
  \put(143,49){$\chi_\varphi^\varphi$}
\end{picture}%
\includegraphics[height=2.9cm]{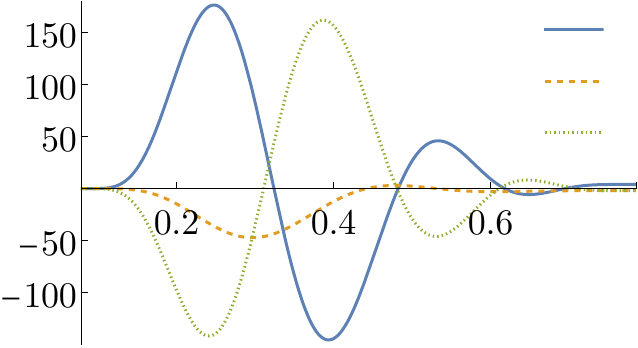}
\caption{ Left: $\nu^\normal$. Middle: $\nu$. Right: curvatures $(\chi_r^r,\chi_\theta^\theta,\chi_\varphi^\varphi)$, for $\theta \in [\theta_1,\theta_2]$.}
\label{figurethree}
\end{figure}

{ 

\paragraph{Axisymmetric formulation.}
 
We denote by $\theta\in[0,\pi]$ the colatitude on the sphere $\Sphe^{2}$, with metric $d\theta^2+(\sin\theta)^2d\varphi^2$, with $\varphi\in\Sphe^{1}$.
 In axisymmetry, $\lambda=\lambda(\theta)$ is a function of $\theta$ supported on an interval $I=(\theta_1,\theta_2)$ and vanishing linearly at the end-points.
In view of $\Deltaslash h = \frac{1}{\sin\theta} \del_\theta\bigl(\sin\theta\del_\theta h\bigr) 
= h'' + \cot\theta \, h'$ (a prime~$'$ denoting a~$\theta$ derivative), we find 
\be\eqalign{
\sigma   \ssA[\nu]
& = 2 \, \bigl(\sigma   ( \nu'' + \cot\theta \nu') \bigr)''
+ 2 \cot\theta \, \bigl(\sigma   ( \nu'' + \cot\theta \nu') \bigr)'
\\
& \quad 
- \cot\theta \bigl( \sigma' \, \nu' \bigr)'
- 4 p(1-p)\big( \sigma'' + \cot\theta \sigma' \big) \, \nu 
\\
& \quad - 4 (1-p) (1+2p) \, \sigma' \, \nu' 
- (5-4p^2) \, \sigma   \big( \nu'' + \cot \theta \, \nu' \bigr),
}
\ee
where $\sigma = \lambda^{2\expoP}$.
This is a fourth-order linear elliptic operator in the variable $\theta\in(\theta_1,\theta_2)$ and is of Fuchsian type at $\theta_1,\theta_2$ with characteristic exponents $2-2\expoP$, $3-2\expoP$, $0$,~$1$, namely has a two-dimensional subset of solutions that remain finite at a given boundary.


\paragraph{Axisymmetric silhouette function.} 

The one-dimensional asymptotic kernel at infinity is determined by solving the equation $\ssA[\nu] = 0$ while imposing boundedness at both boundaries, and imposing any non-vanishing condition, for instance at one boundary. It is convenient to construct first such a $\nu_1$ and next normalize it to get the silhouette $\nu^\normal = 8 \pi \, \nu_1/ \int_I \bigl( \Deltaslash\nu_1 - 4(1-p) \nu_1 ) \sigma \sin \theta d\theta$. 



\paragraph{Gluing Schwarzschild metrics.}

To glue together two Schwarzschild metrics with masses $m_1\geq 0$ (taken to be zero in our test below) and $m_2 \geq 0$, we consider the seed data 
\begin{equation}
\seedg{}_{ij} = \Bigl(1+{1 \over 2r} \seedm(\theta) \Bigr)^4\delta_{ij},
\quad
\seedm(\theta) =  m_1 + (m_2 - m_1) \Sigma(\theta), 
\end{equation}
in which $\seedm: I \to \RR$ provides us with a (local mass) function defined on $S^2$ and $\Sigma(\theta) = \int_{\theta_1}^{\theta} \sigma(\theta') \csc \theta' d\theta'
/ \int_{\theta_1}^{\theta_2} \sigma(\theta') \csc \theta' d\theta'$.
The Hamiltonian constraint reads $\Hcal(\seedg) = - 4r^{-3} \Deltaslash \seedm + \Or(r^{-4})$, so the leading correction to the seed metric is determined by solving $\ssA[\nu] = 4 \, \Deltaslash \seedm$ ---{\it unique up to} a multiple of $\nu^\normal$.


\paragraph{Numerical experiment.} 

We take $p=1/2$ (corresponding to a harmonic exponent equal to $1$ for, both, the metric variable and its dual variable).
From $\nu$, we obtain the metric 
\begin{equation}
\eqalign{
g_{ij} =  \, \delta_{ij} 
+ {1 \over r} \Big( 2\seedm \, \delta_{ij}  
+ \sigma r^3 \del_i \del_j  (\nu/r) - \delta_{ij} \sigma \Deltaslash \nu
\Bigr) 
+ \mathrm{o}\Bigl(\frac{1}{r}\Bigr). 
}
\end{equation}
The leading term in the Ricci curvature $\textbf{Ric}=\chi\, r^{-3}+\mathrm{o}(r^{-3})$ is traceless and diagonal in $(r,\theta,\varphi)$-coordinates, and its components $(\chi_r^r,\chi_\theta^\theta,\chi_\varphi^\varphi)$ are invariant under transformations that preserve the asymptotically { Euclidean} structure.  As an example,
\begin{equation}
\chi_r^r
 = 4 \seedm - 2 \sigma \Deltaslash\nu - \cot\theta (\sigma' \nu')' / 2
+ 2 \sigma' \nu'
+ ((3/2) \Deltaslash\sigma + 4 \sigma ) \, \nu .
\end{equation}
For definiteness, we choose 
$
\lambda(\theta)= 4 (\theta - \theta_1) ( \theta_2 - \theta) / (\theta_2-\theta_1)^2$,
 the interval $[\theta_1,\theta_2]=[\pi/40, \pi/4]\simeq[0.079,0.785]$, and $P=4$, with $m_1=0$ and $m_2=1$. 
The functions $\lambda$ and $\sigma$ are shown in Figure~\ref{figure--3}. 
In Figure~\ref{figurethree}, we display the silhouette $\nu^{\normal}$ (which is close to a constant, as it would be in a non-localized setup), the dual variable $\nu$ (defined modulo a multiple of the silhouette), and the curvatures. 
Observe that $(\chi_r^r,\chi_\theta^\theta,\chi_\varphi^\varphi)$ on the sphere at infinity varies (component-wise) {\it non-monotonically} between  $(0,0,0)$ (Euclidian) and $(4,-2,-2)$ (Schwarzschild).

}

\section{Concluding remarks}
\label{section6} 

\paragraph{Related results.}
 
The construction and analysis of physically relevant solutions to the Einstein constraints is a central topic in the physical, mathematical, and numerical literature; cf.\ the recent reviews by Carlotto~\cite{Carlotto-Review} and Galloway et al.~\cite{GallowayMiaoSchoen}, as well as Henneaux~\cite{Henneaux}. Beig and Chru\'sciel~\cite{BeigChrusciel-2017} studied the gluing problem in linearized gravity. In addition to the variational method, Lichnerowicz and followers proposed the \textit{conformal method}, which we do not attempt to review here~\cite{Gicquaud,Isenberg-1995,IsenbergMaxwellPollack,Maxwell-2021}. In particular, Isenberg~\cite{Isenberg-1995} succeeded to parametrize all closed manifolds representing data sets with constant mean curvature, a result later generalized by Maxwell~\cite{Maxwell-2021}. 


\paragraph{Perspectives.}

Our framework should have interesting applications beyond the localization problem. From a physics viewpoint, the asymptotic kernels { help uncover} the structure of Einstein's constraints at infinity within a (possibly narrow) angular domain.
Furthermore, our construction in~\cite{BLF--PLF} relies on \textit{weighted energy functionals} associated with the localized Hamiltonian and momentum~\eqref{equa:acalew0}.  In combination with elliptic regularity, these enable the desired estimates of the (integral, pointwise) decay of { solutions}.  Beyond their technical use, { they are} interesting concepts in their own right. 



\paragraph*{Acknowledgments. }

The authors were supported by the projects {\it Einstein-PPF}: {\it ``Einstein constraints: past, present, and future''} funded by the Agence Nationale de la Recherche, and {\it Einstein-Waves: \it ``Einstein gravity and nonlinear waves: physical models, numerical simulations, and data analysis''} (101131233), funded by the European Research Council.
 


\end{document}